\documentclass[aps,preprint,showpacs,showkeys,preprintnumbers,amsmath,amssymb]{revtex4}

\usepackage{dcolumn}
\usepackage{mathrsfs}
\usepackage{bm}
\usepackage{amsmath,amssymb,epsfig,float}

\begin{document}

\title{Multipartite entanglement of fermionic systems \\ in noninertial frames}
\author{Jieci Wang and Jiliang Jing\footnote{Corresponding author, Email: jljing@hunnu.edu.cn}}
\affiliation{ Institute of Physics and  Department of Physics,
\\ Hunan Normal University, Changsha, \\ Hunan 410081, P. R. China
\\ and
\\ Key Laboratory of Low-dimensional Quantum Structures
\\ and Quantum
Control of Ministry of Education, \\ Hunan Normal University,
Changsha, Hunan 410081, P. R. China}


\begin{abstract}

The bipartite and tripartite entanglement of a 3-qubit fermionic
system when one or two subsystems accelerated are investigated. It
is shown that all the one-tangles decrease as the acceleration
increases. However, unlike the scalar case, here one-tangles ${\cal
N}_{C_I(AB_I)}$ and  ${\cal N}_{C_I(AB)}$ never reduce to zero for
any acceleration. It is found that the system has only tripartite
entanglement when either one or two subsystems accelerated, which
means that the acceleration doesn't generate bipartite entanglement
and doesn't effect the entanglement structure of the quantum states
in this system. It is of interest to note that the $\pi$-tangle of
the two-observers-accelerated case decreases much quicker than that
of the one-observer-accelerated case and it reduces to a non-zero
minimum in the infinite acceleration limit. Thus we argue that the
qutrit systems are better than qubit systems  to perform  quantum
information processing tasks in noninertial systems.

\end{abstract}

\vspace*{1.5cm}
 \pacs{03.65.Ud, 03.67.Mn, 04.70.Dy}

\keywords{ Multipartite entanglement, Qutrit systems, Fermionic system.}

\maketitle

\section{introduction}

Quantum entanglement is both the central concept and the most
desirable resources for a variety of quantum information processing
tasks \cite{Peres,Boschi, Bouwmeester}, such as quantum
teleportation, super-dense coding, entanglement-based  quantum
cryptography, error correcting codes, and quantum computation. In
the last decade, although many efforts have been made on the study
of the properties of entanglement, the good understanding of such a
resource is only limited in bipartite systems. There's no doubt that
the multipartite entanglement is a valuable physical resource in
large scale quantum information processing and plays an important
role in condensed matter physics. But, in fact, although the
entanglement of multipartite systems can be similarly investigated
as bipartite case, the properties and quantification of entanglement
for higher dimensional systems and multipartite quantum systems are
some issues still to be resolved.

On the other hand, as a combination of general relativity, quantum
field theory and quantum information theory, the relativistic quantum
information has been a focus of research in quantum information
science over recent years
for both conceptual and experimental reasons. In the last few years,
much attention had been given to the study of entanglement shared between
inertial and noninertial observers by discussing how the Unruh or
Hawking effect will influence the degree of entanglement
\cite{Alsing-Mann, Schuller-Mann, Alsing-Milburn, moradi, Qiyuan, jieci1,
jieci2, jieci3, Jason, Ralph, David, Andre, adesso, Eduardo, Juan, Mann2}.
However, it is worth to note that most investigations in the
noninertial system focused on the study of the quantum information
in bipartite systems and only one of the subsystems accelerated.
Fortunately, the tripartite entanglement of
scalar filed between noninertial frames was studied by Mi-Ra Hwang
\emph{et al.} \cite{Hwang} most recently. They showed that the tripartite
entanglement decreases with increase of the acceleration
but different from bipartite entanglement when one
observer moves with an infinite acceleration.

In this paper we will discuss both the bipartite and tripartite entanglement
of Dirac fields in the noninertial frame when one or two observers accelerated.
We are interested in how the accelerations of these observers will
influence the degree of bipartite and tripartite entanglement, and
whether the differences between Fermi-Dirac and Bose-Einstein
statistic will play a role in the entanglement decrease. Our setting
consists of three observers: Alice, Bob and Charlie.
We first assume Alice is in an inertial frame, Bob and Charlie are observing the system from accelerated frames, and then let Alice and Bob stay stationary while
Charlie moves with uniformly accelerations. We consider the Dirac fields, as shown in Refs. \cite{adesso2, E.M, Bruschi}, from the perspective of the observers
who uniformly accelerated, are described as an entangled state
of two modes monochromatic with frequency $\omega_i,\forall i$
\begin{eqnarray}\label{Dirac-vacuum}
|0_{\omega_i}\rangle_{M} =
\cos r_i|0_{\omega_i}\rangle_{I}|0_{\omega_i}\rangle
_{II}+\sin r_i|1_{\omega_i}\rangle_{I}|1_{\omega_i}\rangle
_{II},
\end{eqnarray}
and the only excited state is
\begin{eqnarray}\label{Dirac-excited}
|1_{\omega_i}\rangle_{M}=|1_{\omega_i}
\rangle_{I}|0_{\omega_i}\rangle_{II},
\end{eqnarray}
where $\cos r_i=(e^{-2\pi\omega_i c/a_i}+1)^{-1/2}$, $a_i$ is the
acceleration of the observer $i$.
Considering that an accelerated observer in Rindler region $I$ has
no access to the field modes in the causally disconnected region $II$. By
tracing over the inaccessible modes we will obtain a tripartite state and then
we calculate the tripartite entanglement of the 3-qubit state as well as
bipartite entanglement of all possible bipartite divisions of the tripartite system.

The outline of the paper is as follows. In Sec. II we recall some
measurements of entanglement in quantum information theory, in
particular the negativity and $\pi$-tangle. In Sec. III the bipartite and
tripartite entanglement of Dirac fields when one or two of the observers
accelerated will be discussed. The conclusions are presented
in the last section.

\section{Measures of tripartite entanglement}

It is well known  that there are two remarkable entanglement measures
for a bipartite system $\rho_{\alpha\beta}$,  the concurrence \cite{Coffman}
and the negativity \cite{Vidal}. The former is defined as
\begin{eqnarray}  \label{Concurrence}
C_{\alpha\beta} =\max \left\{ 0,\lambda _{1}-\lambda
_{2}-\lambda _{3}-\lambda _{4}\right\}, \quad\lambda_i\ge
\lambda_{i+1}\ge 0,
\end{eqnarray}
where $\lambda_i$ are the square roots of the eigenvalues of the
matrix $\rho_{\alpha\beta}\tilde{\rho}_{\alpha\beta}$ with $\tilde{\rho}_{\alpha\beta}=(\sigma_y\otimes\sigma_y)\,
\rho_{\alpha\beta}^{*}\,(\sigma_y\otimes\sigma_y)$ is the ``spin-flip" matrix
 and $\sigma_y$ is the Pauli matrix, and the latter is
defined as
\begin{eqnarray}  \label{negativity}
{\cal N}_{\alpha\beta} =\|\rho^{T_\alpha}_{\alpha\beta}\|-1,
\end{eqnarray}
where $T_\alpha$ denotes the partial transpose of $\rho_{\alpha\beta}$ and $\|.\|$ is the
trace norm of a matrix.
Corresponding, there are two entanglement measures which quantify the genuine tripartite
entanglement: three-tangle \cite{Coffman} and $\pi$-tangle \cite{ou}. The three-tangle (or residual tangle), which has many nice properties
but it is a highly difficult problem to compute it analytically except few rare cases, is defined as
\begin{eqnarray}
\tau_{\alpha,\beta,\gamma}=\tau_{\alpha(\beta,\gamma)}-\tau_{\alpha,\beta}-\tau_{\alpha,\gamma},
\end{eqnarray}
where $\tau_{\alpha(\beta,\gamma)}=C^2_{\alpha(\beta,\gamma)}$ and  $\tau_{\alpha,\beta}=C^2_{\alpha,\beta}$.

In order to get a easier calculation here we only adopt the $\pi$-tangle as the quantification of the tripartite entanglement. For any 3-qubit
states $|\Phi\rangle_{\alpha \beta\gamma}$, the entanglement
quantified by the negativity between $\alpha$ and $\beta$, between $\alpha$ and
$\gamma$, and between $\alpha$ and the overall subsystem $\beta \gamma$ satisfies the
following Coffman-Kundu-Wootters (CKW) monogamy inequality \cite{Coffman}
\begin{equation}  \label{g1}
{\cal N}^2_{\alpha\beta}+{\cal N}_{\alpha\gamma}^2\leq {\cal N}^2_{\alpha(\beta\gamma)},
\end{equation}
where ${\cal N}_{\alpha\beta}$ is `two-tangle', which is the negativity of the mixed state
$\rho_{\alpha \beta}=\mathrm{Tr}_{\gamma}(|\Phi\rangle_{\alpha \beta\gamma}\langle\Phi|)$
and ${\cal N}^2_{\alpha(\beta\gamma)}$ is `one-tangle', defined as
${\cal N}_{\alpha(\beta\gamma)} =\|\rho^{T_\alpha}_{\alpha\beta\gamma}\|-1$ .
The difference between the two sides of Eq.(\ref{g1}%
) can be interpreted as the residual entanglement
\begin{equation}  \label{h1}
\pi_{\alpha}= {\cal N}^2_{\alpha(\beta\gamma)}-{\cal N}^2_{\alpha \beta}-{\cal N}_{\alpha\gamma}^2.
\end{equation}
Likewise, we have
\begin{equation}  \label{h2}
\pi_{\beta}= {\cal N}^2_{\beta(\alpha\gamma)}-{\cal N}^2_{\beta\alpha}-{\cal N}_{\beta\gamma}^2,
\end{equation}
and
\begin{equation}  \label{h3}
\pi_{\gamma}= {\cal N}^2_{\gamma(\alpha \beta)}-{\cal N}^2_{\gamma \alpha}-{\cal N}_{\gamma \beta}^2.
\end{equation}
The $\pi$-tangle $\pi_{\alpha\beta\gamma}$ is defined as the average
of $\pi_{\alpha}$, $\pi_{\beta}$, and $\pi_{\gamma}$, i.e.,
\begin{equation}  \label{ptangle}
\pi_{\alpha\beta\gamma}=\frac{1}{3}(\pi_{\alpha}+\pi_{\beta}+\pi_{\gamma}).
\end{equation}

\section{Behaviors of tripartite entanglement when one or two observer accelerated}

 We consider a tripartite system which is consist of three subsystems, name Alice as the observer of the first part of the system, Bob and Charlie are the observers of the second  and third parts respectively. They shared
a GHZ state
\begin{eqnarray}\label{initial}
|\Phi\rangle_{ABC}=\frac{1}{\sqrt{2}}(|0\rangle_{A}|0\rangle_{B}|0\rangle_{C}
+|1\rangle_{A}|1\rangle_{B}|1\rangle_{C}),
\end{eqnarray}
where $|0\rangle_{A(B,C)}$ $|1\rangle_{A(B,C)}$ are vacuum states and
from an inertial observer. Using Eqs. (\ref{negativity}) and (\ref{ptangle}),
we can easily get
\begin{eqnarray} \label{otp}
\nonumber&&{\cal N}_{A(BC)}={\cal N}_{B(AC)}={\cal N}_{C(AB)}=1,\\
\nonumber&&{\cal N}_{AB}={\cal N}_{BC}={\cal N}_{CA}=0,\\
\nonumber&&\pi_{ABC}=1,
\end{eqnarray}
where ${\cal N}_{A(BC)}$, ${\cal N}_{AB}$ and
$\pi_{ABC}$ are the `one-tangle', `two-tangle' and `$\pi$-tangle' of
 state (\ref{initial}) from a inertial viewpoint.
Then we let Alice stays stationary while Bob and Charlie move
with uniform accelerations. Since Bob
and Charlie are accelerated, we should map
the second and third partition of this state into the Rindler Fock
space basis. Using  Eqs.
(\ref{Dirac-vacuum}) and (\ref{Dirac-excited}) we can rewrite Eq.
(\ref{initial}) in terms of Minkowski modes for Alice and Rindler
modes for Bob and Charlie
\begin{eqnarray}  \label{state}
\nonumber|\Phi\rangle_{AB_IC_I}&=&\frac{1}{\sqrt{2}}\bigg[ \cos r_b
\cos r_c|0\rangle_{A}|0\rangle_{B_I}|0\rangle_{B_{II}}|0\rangle_{C_I}
|0\rangle_{C_{II}}+\cos r_b \sin r_c|0\rangle_{A}|0\rangle_{B_I}
|0\rangle_{B_{II}}|1\rangle_{C_I}|1\rangle_{C_{II}}\\&& \nonumber
+\cos r_b \sin r_c|0\rangle_{A}|1\rangle_{B_I}|1\rangle_{B_{II}}
|0\rangle_{C_I}|0\rangle_{C_{II}}+\sin r_b \sin r_c|0\rangle_{A}
|1\rangle_{B_I}|1\rangle_{B_{II}}|1\rangle_{C_I}|1\rangle_{C_{II}}
\\&&+|1\rangle_{A}|1\rangle_{B_I}|0\rangle_{B_{II}}|1\rangle_{C_I}
|0\rangle_{C_{II}}\bigg].
\end{eqnarray}

Let us first calculate the one-tangle between subsystem $A$ and the
overall subsystem $B_IC_I$ by use of Eq. (\ref{negativity}). Tracing
over the inaccessible modes $B_{II}$ and $C_{II}$ we obtain a
density matrix
\begin{eqnarray}  \label{eq:state}
\nonumber\rho_{AB_IC_I}&=&\frac{1}{2}\bigg[\cos^2 r_b \cos^2 r_c|000\rangle\langle000|
+\cos^2 r_b \sin^2 r_c|001\rangle\langle001|\\&&\nonumber+\sin^2 r_b \cos^2 r_c|010
\rangle\langle010|+\sin^2 r_b \sin^2 r_c|011\rangle\langle011|\\&&+
\cos r_b \cos r_c(|111\rangle\langle000|+|000\rangle\langle111|+|111\rangle
\langle111|\bigg],
\end{eqnarray}
where $|lmn\rangle=|l\rangle_{A}|m\rangle_{B_{I}}|n\rangle_{C_{I}}$.
Then we can easily get the partial transpose subsystem $A$ of Eq. (\ref{eq:state})
\begin{eqnarray}  \label{ta}
\nonumber\rho^{T_A}_{AB_IC_I}&=&\frac{1}{2}\bigg[\cos^2 r_b \cos^2 r_c|000
\rangle\langle000|+\cos^2 r_b \sin^2 r_c|001\rangle\langle001|\\&&
+\nonumber\sin^2 r_b \cos^2 r_c|010\rangle\langle010|+\sin^2 r_b
 \sin^2 r_c|011\rangle\langle011|\\&&+
\cos r_b \cos r_c(|011\rangle\langle100|+|100\rangle\langle011|+|111
\rangle\langle111|  \bigg],
\end{eqnarray}
from which we can get $(\rho^{T_A}_{AB_IC_I})^{\dag}$ and the  negativity
${\cal N}_{A(B_IC_I)}$ is found to be
\begin{eqnarray}\label{N1}
\nonumber{\cal N}_{A(B_IC_I)}&=&\frac{1}{2}\bigg[\cos r_b \cos r_c+
\cos^2 r_c+\cos^2 r_b \sin^2 r_c\\&& +\sqrt{\cos^2 r_b \cos^2 r_c
+\sin^4 r_b \sin^4 r_c} -1\bigg]
\end{eqnarray}

Similarly, we can also get
\begin{eqnarray}\label{N2}
\nonumber{\cal N}_{B_I(AC_I)}&=&\frac{1}{2}\bigg[\cos r_b \cos r_c
+\cos^2 r_b+\sin^2 r_b \sin^2 r_c\\&&+\cos r_c\sqrt{\cos^2 r_b
+\sin^4 r_b \cos^2 r_c} -1\bigg],
\end{eqnarray}
and
\begin{eqnarray}\label{N3}
\nonumber{\cal N}_{C_I(AB_I)}&=&\frac{1}{2}\bigg[\cos r_b \cos r_c+\sin^2 r_b
+\cos^2 r_b \cos^2 r_c\\&&+\cos r_b\sqrt{\cos^2 r_c +\sin^4 r_c \cos^2 r_b} -1\bigg].
\end{eqnarray}

\begin{figure}[ht]
\includegraphics[scale=0.8]{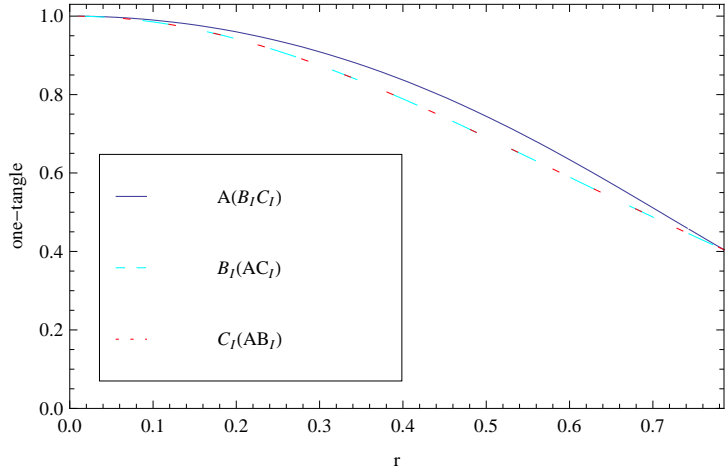}
\caption{\label{GHZonetABI2}(Color online) The negativity ${\cal N}_{A(B_IC_I)}$(solid line),
${\cal N}_{B_I(AC_I)}$ (dashed line),
and ${\cal N}_{C_I(AB_I)}$ (dotted line)
of two-observers-accelerated case as a function of the acceleration parameter $r=r_b=r_c$.}
\end{figure}

The properties of all the one-tangles of $\rho_{AB_IC_I}$ are shown
in Fig. \ref{GHZonetABI2} with $r_b=r_c=r$. It is shown that all the one-tangles
equal to one when $r=0$, which is exactly the value of one-tangles
in Eq. (\ref{initial}) obtained in the inertial frame. All of them
decrease as the accelerations of Bob and Charlie increase, which
is similar to the behaviors of bipartite entanglement of Dirac field
\cite{Alsing-Mann} and tripartite one-tangle
of scalar field when one of the observers accelerated \cite{Hwang}.
Note that ${\cal N}_{B_I(AC_I)}={\cal N}_{C_I(AB_I)}$ for all
accelerations which indicate Bob and Charlie's subsystems
is symmetry in this case. It is worthwhile to note
that unlike the scalar case the one-tangle $N_{C_I(AB)}$ goes to zero
when Charlie moves with infinite acceleration, here $N_{C_I(AB_I)}$ and  $N_{C_I(AB)}$
never goes to zero for any acceleration.  We argue that this different is
due to the differences between Fermi-Dirac and Bose-Einstein statistics \cite{Eduardo} rather than because the observers cannot
access to the entanglement of the subsystems who
moves with infinite acceleration respect to them, as the authors stated
in Ref. \cite{Hwang}. What's surprising is that in the case of
 both Bob and Charlie move with infinite accelerations ($r=\pi/4$),
${\cal N}_{A(B_IC_I)}={\cal N}_{B_I(AC_I)}
={\cal N}_{C_I(AB_I)}=\frac{1-\sqrt{5}}{8}$, which means that
there is no difference between all the subsystems
$A$, $B_I$ and $C_I$ in this limit.

Now, let us compute the two-tangle between subsystems $A$ and $B_I$,
tracing the qubit of subsystem $C_I$ we obtain
\begin{eqnarray}
\nonumber{\cal \rho}_{AB_I}=\frac{1}{2}\left(
                                \begin{array}{cccc}
                                  \cos r_b^2\cos^2 r_c + \cos r_b^2 \sin^2 r_c & 0 & 0 & 0 \\
                                  0 & \sin^2 r_b \cos^2 r_c + \sin r_b^2 \sin^2 r_c & 0 & 0 \\
                                  0 & 0 & 0 & 0 \\
                                  0 & 0 & 0 & 1 \\
                                \end{array}
                              \right).
\end{eqnarray}
Using this matrix and Eq.(\ref{negativity}) we can obtain the negativity
${\cal N}_{AB_I}=0$, which means there is no bipartite entanglement
between mode $A$ and $B_I$  in spite of Bob and Charlie
with accelerations. Similarly, it is found that
${\cal N}_{AC_I}={\cal N}_{B_IC_I}=0$.
Note that the CKW inequality  \cite{Coffman},
${\cal N}^2_{\alpha\beta}+{\cal N}_{\alpha\gamma}^2
\leq {\cal N}^2_{\alpha(\beta\gamma)}$, is saturated
for any acceleration parameter $r$.

Then by use of Eqs. (\ref{h1}$\sim$\ref{ptangle}), the $\pi$-tangle of
our system is found to be
\begin{eqnarray}
\pi_{AB_IC_I}=\frac{1}{3}(\pi_{A}+\pi_{B_I}+\pi_{C_I})
=\frac{1}{3}\bigg[{\cal N}^2_{A(B_IC_I)}+{\cal N}^2_{B_I(AC_I)}
+{\cal N}^2_{C_I(AB_I)}\bigg],
\end{eqnarray}
where ${\cal N}_{A(B_IC_I)}$, ${\cal N}_{B_I(AC_I)}$
and ${\cal N}_{C_I(AB_I)})$ are given by Eq.(\ref{N1}$\sim$\ref{N3}) respectively.

In order to better understand the multipartite entanglement
in the noninertial frames, we also compute the entanglement of a tripartite system include two inertial subsystems and one noninertial subsystem, i.e.,
let Alice and Bob stay stationary and Charlie moves with uniform
acceleration. They shared the same GHZ state Eq. (\ref{initial}) at the
same point in the Minkowski spacetime. According to the preceding calculations,
we can obtain
\begin{eqnarray} \label{otp1}
\nonumber&&{\cal N}_{A(BC_I)}={\cal N}_{B(AC_I)}=\cos r_c,\\
\nonumber&&{\cal N}_{C_I(AB)}=\frac{1}{2}(\cos r_c+\cos^2 r_c
+\sqrt{\cos^2 r_c+\sin^4 r_c}-1),\\
&&{\cal N}_{AB}={\cal N}_{BC_I}={\cal N}_{C_IA}=0,
\end{eqnarray}
where ${\cal N}_{A(BC_I)}$ and ${\cal N}_{AB}$ are the one-tangle
and two-tangle of the one-observer-accelerated case. It is worthy to
notice that the CKW inequality is also saturated for any
acceleration parameter $r$ in this case. From these facts we arrive
at the conclusion that this inequality is valid in both  inertial
and noninertial frames.

\begin{figure}[ht]
\includegraphics[scale=0.8]{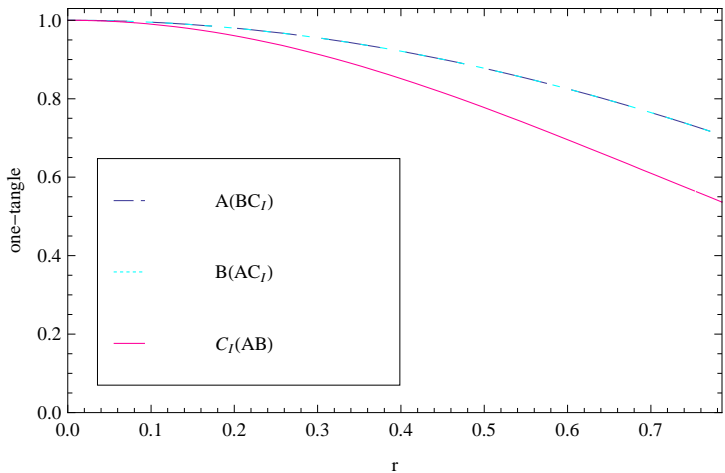}
\caption{\label{GHZonetABI}(Color online) The one-tangles ${\cal N}_{A(BC_I)}$( dashed line),
${\cal N}_{B(AC_I)}$ (dotted line),
and ${\cal N}_{C_I(AB)}$ (solid  line)
of one-observer-accelerated case as a function of the acceleration parameter $r=r_c$.}
\end{figure}

We plot the one-tangles of this case in  Fig. \ref{GHZonetABI} and
find that: (i) all of them decrease as the acceleration of Charlie
increases; (ii) ${\cal N}_{A(BC_I)}={\cal N}_{B(AC_I)}$ for all
accelerations; (iii) the one-tangle $N_{C_I(AB)}$ never goes to zero
for any acceleration. However, it is interesting to note that in
this case ${\cal N}_{A(BC_I)}={\cal N}_{B(AC_I)} \neq{\cal
N}_{C_I(AB)}$ when Charlie moves with infinite acceleration, which
is very different from the two-observers-accelerated case. We are
not sure whether this is an individual case only appears in the
fermionic systems because it probably related to the incomplete
definition of the one-tangle in the noninertial frames.   Thus, it
seems to be interesting to repeat the calculation of this paper for
other systems and by making use of other entanglement measurements.
It is shown again that all the two-tangles equal to zero in this
case, which is exactly the same as the two-tangles obtained in the
inertial frame. That is to say, either one or two subsystems of the
tripartite state are accelerated, there is no bipartite entanglement
in this system. The acceleration doesn't generate bipartite
entanglement and the entanglement structure of the quantum state
doesn't changed. It is interesting to note that in Ref.
\cite{Alsing-Mann}, there was no tripartite entanglement between
observers Alice, Rob and Anti-Rob, i.e, all the entanglement is
bipartite when one of the observers static and the other
accelerated. However, here we find that there is no bipartite
entanglement, all the entanglement of this system is in form of
tripartite entanglement.

\begin{figure}[ht]
\includegraphics[scale=0.8]{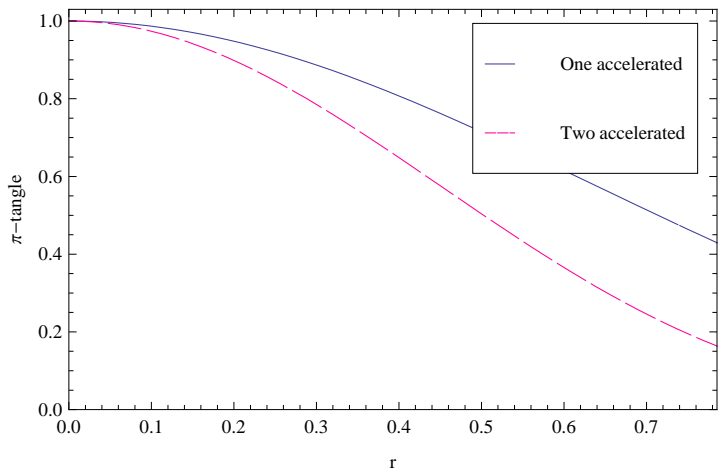}
\caption{\label{GHZonetpare}(Color online) The $\pi$-tangle
$\pi_{ABC_I}$ of one-observer-accelerated case (solid line) versus
$\pi_{AB_IC_I}$ of two-observers-accelerated case (dashed line),
as a function of the acceleration parameter $r=r_b=r_c$.}
\end{figure}

By use of Eq. (\ref{otp}) we get the $\pi$-tangle of
the one-observer-accelerated system $
\pi_{ABC_I}=\frac{{\cal N}^2_{A(BC_I)}+{\cal N}^2_{B(AC_I)}
+{\cal N}^2_{C_I(AB)}}{3}$. For comparison, we plot the $\pi_{ABC_I}$ of
this case and the $\pi_{AB_IC_I}$ of  two-observers-accelerated case
in Fig. \ref{GHZonetpare}.

In Fig. (\ref{GHZonetpare})  we plot the $\pi$-tangle $\pi_{ABC_I}$
of the one-observer-accelerated case and $\pi_{AB_IC_I}$ of the
two-observers-accelerated case as a function of the acceleration
parameter $r=r_b=r_c$, which shows how the acceleration changes
tripartite entanglement. In the case of zero acceleration,
$\pi_{ABC_I}=\pi_{AB_IC_I}=1$.  With the increasing of
accelerations, the $\pi$-tangles decrease monotonously for both of
these two cases. It is shown that the decrease speed of the
tripartite entanglement of two-observers-accelerated case is much
quicker than that of one-observer-accelerated case as expected.
Recall that the loss of entanglement was explained as the
information formed in the inertial system was leaked into the
causally disconnected region \cite{adesso,E.M,jieci2} due to the
Unruh effect. The quicker decreases of entanglement in
two-observer-accelerated case is attribute to the information both
in Bob and Charlie's subsystems are redistributed into the
unaccessible regions as the growth of acceleration. In the limit of
infinite acceleration, the tripartite entanglement of
two-observers-accelerated case reduces to a lower minimum but never
vanishes. It is interesting to note that either the scalar system or
Dirac system and either one or two observers are accelerated, the
tripartite entanglement never vanishes for any acceleration. Thus we
can arrive a striking conclusion that the quantum entanglement in
tripartite system is a better resource to perform  quantum
information processing  such as teleportation. We can also perform
such quantum information tasks by use of the tripartite entanglement
when some observers are falling into a black hole while others are
hovering outside the event horizon.

\section{summary}

The effect of acceleration on bipartite and tripartite entanglement
of a 3-qubit Dirac system when one or two subsystems accelerated is
investigated. It is shown that all the one-tangles decrease as the
accelerations of Bob and Charlie increase. However, unlike the
scalar case in which the one-tangle ${\cal N}_{C_I(AB)}$ goes to
zero when Charlie moves with infinite acceleration, here ${\cal
N}_{C_I(AB_I)}$ and ${\cal N}_{C_I(AB)}$ never reduce to zero for
any acceleration. It is also shown that the CKW inequality is valid
in the noninertial systems. It is of interest to note that ${\cal
N}_{A(B_IC_I)}= {\cal N}_{B_I(AC_I)}={\cal N}_{C_I(AB_I)}$ in the
infinite acceleration limit, which means that there is no difference
between all the subsystem $A$, $B_I$ and $C_I$ in this limit. It is
found that either one or two subsystems of the tripartite state
accelerated, there is no bipartite entanglement in this system,
i.e., all entanglement of this system is in form of tripartite
entanglement. The acceleration doesn't generate bipartite
entanglement in this system and doesn't change the entanglement
structure of the quantum state. It is also found that the
$\pi$-tangle of the two-observers-accelerated case decreases  much
quicker than that of the one-observer-accelerated case and it
reduces to a non-zero minimum in the infinite acceleration limit. It
is worthy to mention that both for scalar \cite{Hwang} and Dirac
fields, and either one or two observers are accelerated, the
tripartite  entanglement doesn't vanish for any acceleration. That
is to say, the quantum entanglement in tripartite system is a better
resource than bipartite entanglement to perform  quantum information
processing such as teleportation. We can also perform such quantum
information tasks by use of tripartite entanglement when one or two
observers are falling into a black hole while others hovers outside
the event horizon. The discussions of this paper can be also applied
to the investigations of multipartite entanglement and quantum
correlations in the curved spacetime \cite{Qiyuan, E.M, jieci1} as
well as the properties of multipartite Gaussian entanglement in
noninertial frames \cite{adesso}. Therefore, further investigation
by using the results in this paper will not only help us deeply
understand genuine multipartite entanglement but also helpful to
give a better insight into the entanglement entropy and information
paradox of black holes.

\begin{acknowledgments}

This work was supported by the National
Natural Science Foundation of China under Grant No 10875040;  a key
project of the National Natural Science Foundation of China under
Grant No 10935013;  the National Basic Research of China under Grant
No. 2010CB833004,  PCSIRT under Grant No.
IRT0964, and the Construct Program  of the National Key Discipline.

\end{acknowledgments}

\end{document}